\documentclass[aps,prl,twocolumn,reprint,superscriptaddress]{revtex4-2}
\usepackage{graphicx}
\usepackage{amsmath}
\usepackage{amssymb}
\usepackage{colordvi}
\usepackage{mathrsfs}
\usepackage{bm}
\usepackage{verbatim}
\usepackage{dcolumn}
\usepackage{epsfig}
\usepackage{subfigure}
\usepackage[colorlinks,allcolors=blue]{hyperref}
\usepackage{ulem}
\usepackage{dsfont}

\begin{document}
\title{An ab-initio study on engineering quantum anomalous Hall effect in compensated antiferromagnet MnBi$_{2}$Te$_{4}$}
\author{Zeyu Li}
\affiliation{International Centre for Quantum Design of Functional Materials, CAS Key Laboratory of Strongly-Coupled Quantum Matter Physics, and Department of Physics, University of Science and Technology of China, Hefei, Anhui 230026, China}
\affiliation{Hefei National Laboratory, University of Science and Technology of China, Hefei 230088, China}
\author{Yulei Han}
\affiliation{Department of Physics, Fuzhou University, Fuzhou, Fujian 350108, China}
\author{Zhenhua Qiao}
\email[Correspondence author:~~]{qiao@ustc.edu.cn}
\affiliation{International Centre for Quantum Design of Functional Materials, CAS Key Laboratory of Strongly-Coupled Quantum Matter Physics, and Department of Physics, University of Science and Technology of China, Hefei, Anhui 230026, China}
\affiliation{Hefei National Laboratory, University of Science and Technology of China, Hefei 230088, China}
\date{\today}

\begin{abstract}
Recently, the quantum anomalous Hall effect (QAHE) has been theoretically proposed in compensated antiferromagnetic systems by using the magnetic topological insulator model [see arXiv:2404.13305 (2024)]. However, the related and systematic study based on a realistic material system is still limited. As the only experimentally realized antiferromagnetic topological insulator, MnBi$_{2}$Te$_{4}$ becomes a vital platform for exploring various topological states. In this work, by using the comprehensive first-principles calculations, we demonstrate that the QAHE can also be realized in compensated antiferromagnetic even-septuple-layer MnBi$_{2}$Te$_{4}$ without combined parity-time ($\mathcal{PT}$) symmetry. Using a magnetic topological insulator model, the layer-resolved Chern number is calculated to further understand the physical origin of different Chern numbers. The application of external hydrostatic pressure can strengthen the Te-Te quasicovalent bond due to the dramatic compression of the van der Waals gap. Thus, the resulting topological nontrivial gap can exceed the room-temperature energy scale in a wide range of pressures. Additionally, we find that constructing MnBi$_{2}$Te$_{4}$/CrI$_{3}$ heterostructure can realize the compensated antiferromagnetic configurations with QAHE. Our findings illustrate the realization of QAHE in compensated antiferromagnetic even-septuple-layer MnBi$_{2}$Te$_{4}$ and provide a reliable strategy to obtain the corresponding magnetic configurations.
\end{abstract}

\maketitle

\section{\uppercase\expandafter{\romannumeral1}. Introduction} 
The quantum anomalous Hall effect (QAHE) possesses dissipation-free chiral edge states that have long been sought for exploring novel quantum effects and developing next-generation low-power-consumption electronic devices~\cite{QAHEHaldane,half-q}. To acquire the platforms that can achieve the QAHE, tremendous efforts have been made. Up to now, the QAHE has been observed in magnetic topological insulators~\cite{Cr-Bi2Se3,Exp-Cr-Bi2Se3,Rashba-Graphene,2dTI}, twisted bilayer graphene~\cite{TBG} or WSe$_{2}$/MoTe$_{2}$~\cite{TB-TMD}, and rhombohedral graphene~\cite{rhombo-gra}. In these systems, the QAHE is the product of topological bands and broken time-reversal symmetry $\mathcal{T}$. Usually, $\mathcal{T}$ is broken by ferromagnetism induced by magnetic atoms~\cite{mag-atom} or strongly correlated electrons~\cite{e-e}. As another magnetic phase, antiferromagnetism is also able to break time-reversal symmetry. Thus, investigating the QAHE in antiferromagnets makes sense. Nonetheless, the synergistic effect of sublattice-transposing symmetries and time-reversal symmetry exists in the majority of antiferromagnets, giving rise to zero Berry curvature in the first Brillouin zone~\cite{alter1}. Recently, several works have indicated that the QAHE can be realized in antiferromagnet whose sublattice-transposing symmetries are removed~\cite{alter2,alqahe1,alqahe2,alqahe3}. Antiferromagnets are anticipated to play a significant role in studying various topological states since they are more abundant in nature and uneasy prone to external fields than ferromagnets.

In the past five years, the successful synthesis of intrinsic magnetic topological insulator MnBi$_{2}$Te$_{4}$ has provided an ideal playground to investigate the interplay between topological states and magnetism. Antiferromagnetic bulk MnBi$_{2}$Te$_{4}$ is a topological insulator protected by combined sublattice-transposing and time-reversal symmetry~\cite{MBT-TI1,MBT-TI2}. Deviating the protected surface, such as (111) surface, the topological axion state emerges~\cite{MBT-axion}. Type II magnetic Weyl semimetal can also be realized when the interlayer coupling of bulk MnBi$_{2}$Te$_{4}$ becomes ferromagnetic by applying an external magnetic field~\cite{Weyl-MnBiTe}. Because MnBi$_{2}$Te$_{4}$ is bonded by the van der Waals gap, it is easy to exfoliate few-layer samples with intriguing even-odd layer-dependent physics~\cite{layer-dep1,layer-dep2,layer-dep3,layer-dep4}. For the even-septuple-layer case, the magnetizations of the top and bottom layers are antiparallel, harboring topological axion states~\cite{Weyl-MnBiTe}. In contrast, for odd-septuple-layer case, the magnetizations of the top and bottom layers are parallel generating QAH states~\cite{QAHE-MnBiTe}. Consequently, these studies suggest that the magnetic configurations of MnBi$_{2}$Te$_{4}$ are closely related to its various topological states. Many experimental efforts have been made to manipulate the magnetic phases of MnBi$_{2}$Te$_{4}$~\cite{MBT1CrI3,MBT2CrI3}. Given the relatively weak interlayer interaction, some special magnetic configurations beyond A-type antiferromagnetic may be realized. Several works have proved that field training methods can yield them~\cite{mag-train1,mag-train2}, which enables further exploration of the interaction between topological states and magnetism~\cite{QAHE-c1}. \textcolor{red}{According to the model calculations of our earlier work~\cite{QAHE-c2}, the QAHE can be realized in MnBi$_{2}$Te$_{4}$ film with a compensated antiferromagnetic configuration. To thoroughly investigate the conditions and scheme of realization, it is eager to conduct first-principles calculations.}

In this work, we systematically study the topological properties of even-septuple-layer MnBi$_{2}$Te$_{4}$ beyond A-type antiferromagnetic configuration. Our first-principle calculations show that the QAHE can be realized in even-septuple-layer MnBi$_{2}$Te$_{4}$ with compensated antiferromagnetic configurations whose magnetic moments of the outermost Mn$^{2+}$ ions are aligned in the same direction. The topologically nontrivial gaps range from 7 meV to 15 meV. Using the tight-binding model, we calculate the layer-resolved Chern number, revealing the distribution of Chern number within this multilayer system. When a moderate external hydrostatic pressure is applied, the topologically nontrivial gaps increase significantly due to the enhanced interlayer coupling strength. Then, we confirm the possibility of realizing the compensated antiferromagnetic configurations by building CrI$_{3}$/MnBi$_{2}$Te$_{4}$ heterostructure. Our work demonstrates that even-septuple-layer MnBi$_{2}$Te$_{4}$ is a superior platform for exploring topological states with compensated antiferromagnetic orders.

\begin{figure}[t]
  \centering
  \includegraphics[width=0.48\textwidth]{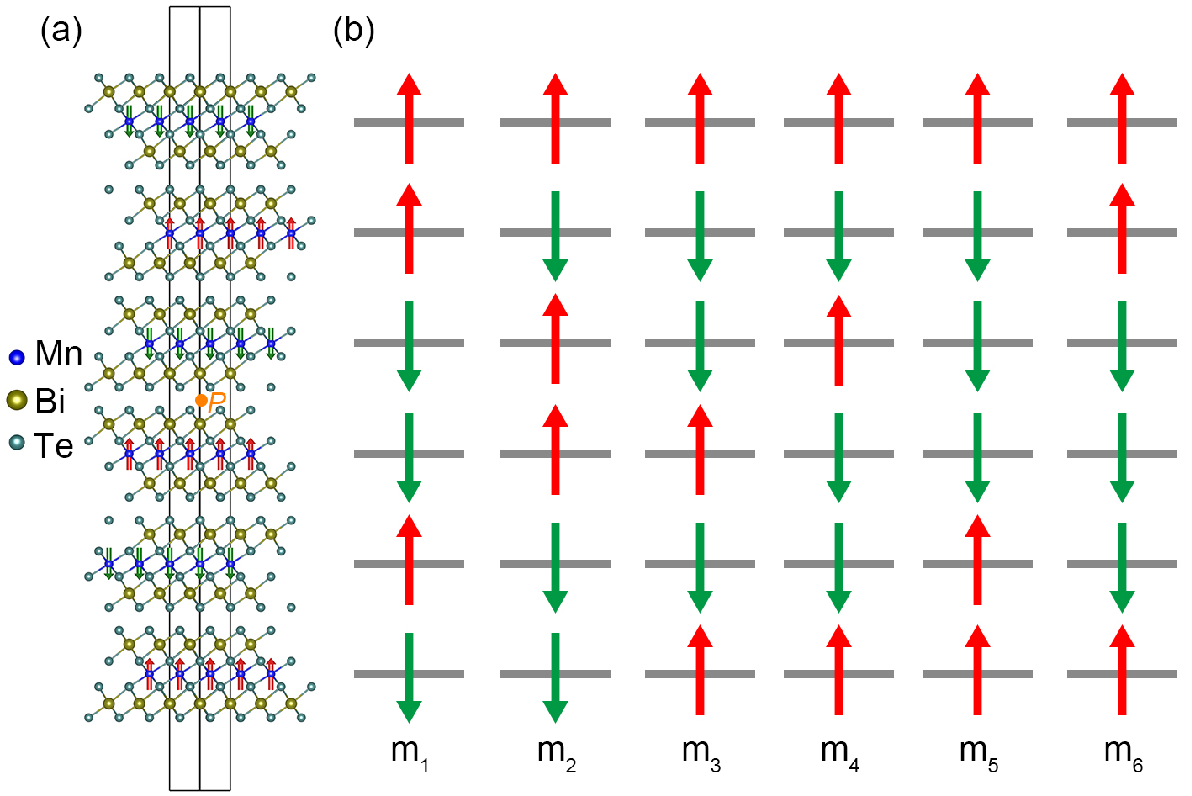}
  \caption{(a) Side view of six-septuple-layer MnBi$_{2}$Te$_{4}$ with A-type antiferromagnetic order, respectively. (b) The schematic diagram of the compensated antiferromagnetic orders without $\mathcal{PT}$ symmetry for six-septuple-layer MnBi$_{2}$Te$_{4}$.}
  \label{fig1}
\end{figure}
\begin{figure*}[t]
	\centering
	\includegraphics[width=0.95\textwidth]{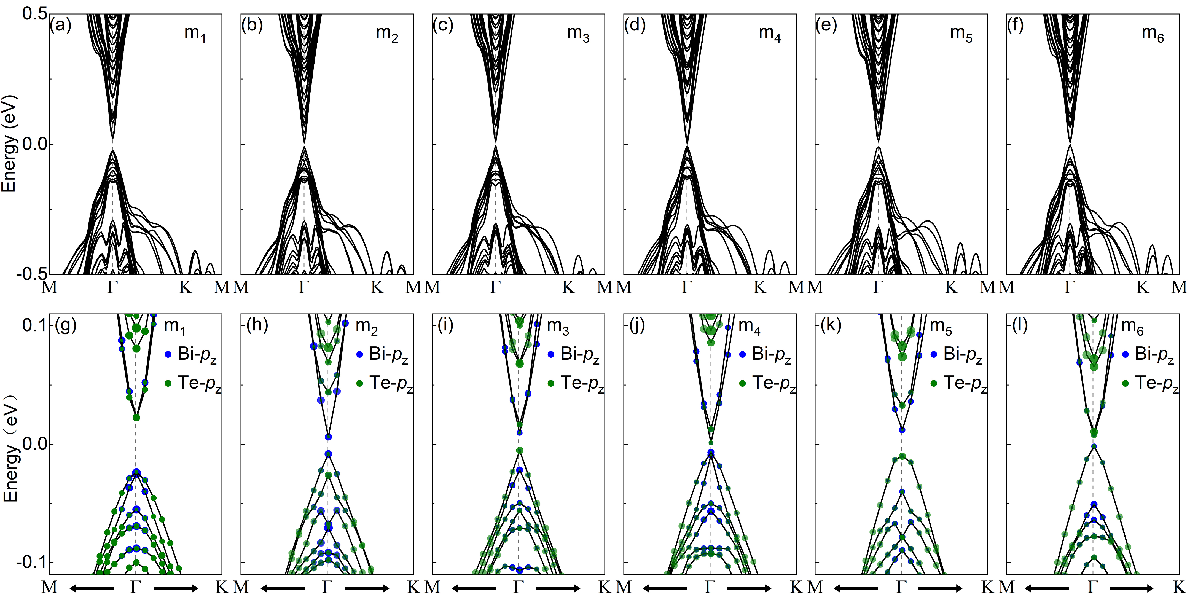}
	\caption{(a)-(f) The band structures of six-septuple-layer MnBi$_{2}$Te$_{4}$ with six compensated antiferromagnetic configurations in the presence of spin-orbit coupling, respectively. (g)-(l) Corresponding zooming-in orbital projected band structures around the $\Gamma$ point in the presence of spin-orbit coupling, respectively. The blue and green colors signify the Bi-$p_{z}$ and Te-$p_{z}$ orbital components, respectively.}
	\label{fig2}
\end{figure*}
\begin{figure}[t]
	\centering
	\includegraphics[width=0.48\textwidth]{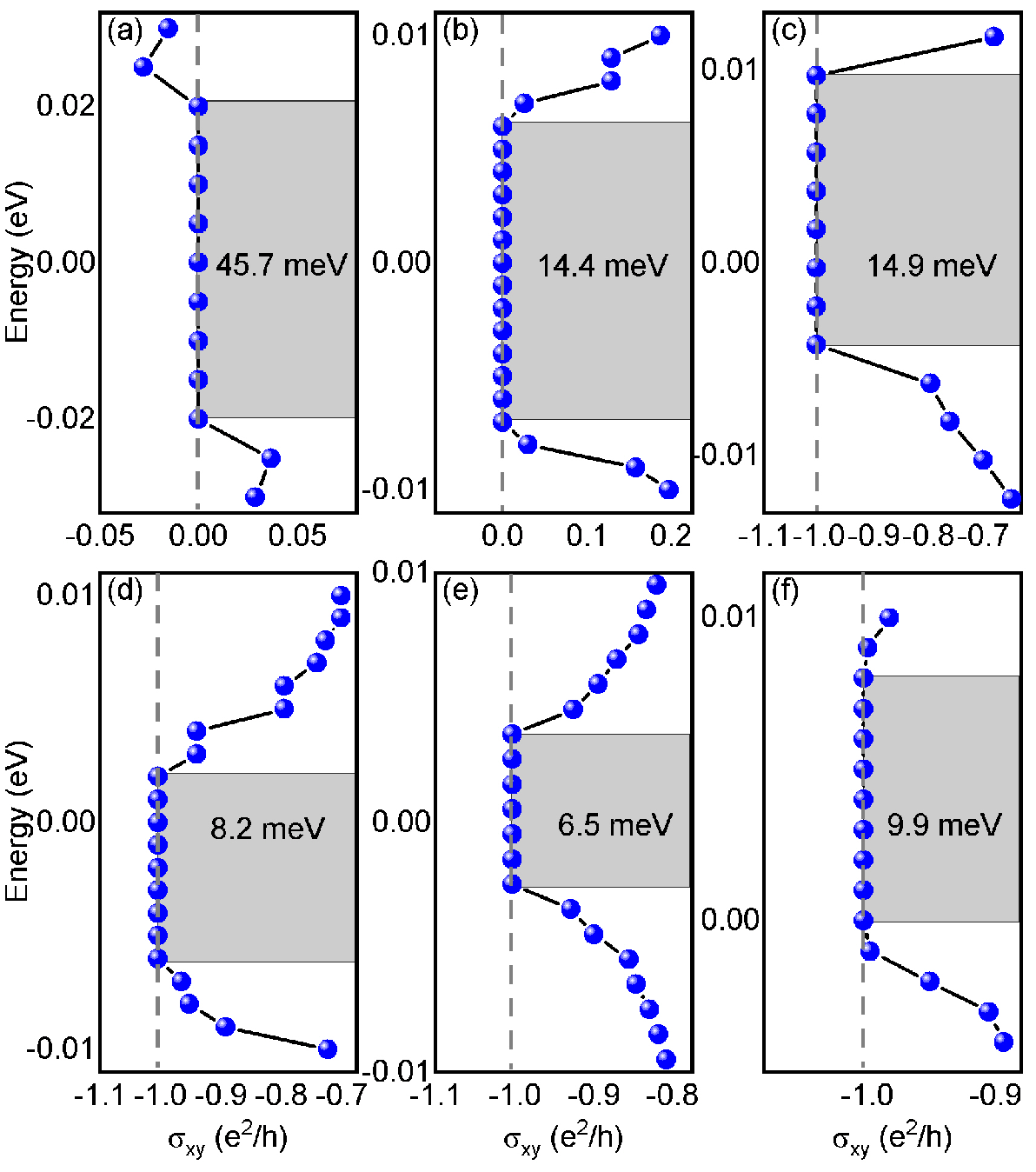}
	\caption{(a)-(f) The anomalous Hall conductivity of m$_{1}$, m$_{2}$, m$_{3}$, m$_{4}$, m$_{5}$, and m$_{6}$ magnetic configurations, respectively. Zero Hall plateaus are observed in m$_{1}$ and m$_{2}$ magnetic configurations, respectively. While the quantized Hall conductance plateaus with $\mathcal{C}=-1$ are observed in m$_{3}$, m$_{4}$, m$_{5}$, and m$_{6}$ magnetic configurations, respectively.}
	\label{fig3}
\end{figure}

\section{\uppercase\expandafter{\romannumeral2}. Calculation Methods} 
Our calculations were performed using the projected-augmented-wave method as implemented in the VASP package~\cite{vasp}, and the generalized gradient approximation exchange-correlation~\cite{GGA} potential was used. The kinetic cutoff energy of plane wave was set to be 500 eV. The Brillouin zone was sampled with a Gamma-centered 12$\times$12$\times$1 grid based on the scheme proposed by Monkhorst-Pack for the calculation of structural optimization and electronic structures~\cite{MP}. A vacuum buffer space over 18~{\AA} was included to prevent interaction between adjacent slabs. The convergence criterion was set to be 10$^{-5}$ eV and 10$^{-7}$ eV for energy in optimization and self-consistent field calculations, respectively. During structural optimization, all atoms were fully relaxed, and forces were converged to less than 0.01 eV/{\AA}. For the Mn and Cr atoms, the GGA+U method was used with the on-site repulsion energy U = 5.00 and 3.00 eV\textcolor{red}{~\cite{GGA-U}},respectively. The interlayer van der Waals interaction is described by DFT-D3 method. The structural evolution of bulk MnBi$_{2}$Te$_{4}$ under various pressures up to 9 GPa is calculated. The corresponding structural parameters of bulk MnBi$_{2}$Te$_{4}$ are applied to four- and six-septuple-layer MnBi$_{2}$Te$_{4}$, respectively. The maximally localized Wannier functions were constructed by using the software package WANNIER90 with interfacing VASP~\cite{wannier90,wannier90-QAHE}. The anomalous Hall conductivity was obtained by the interpolation of maximally localized Wannier functions. The anomalous Hall conductivity was obtained by summing Berry curvatures over all occupied valence bands:
\begin{equation}
\sigma_{\alpha\beta}=-\frac{e^2}{\hbar}\int_{BZ}\frac{dk}{(2\pi)^3}\begin{matrix}\sum_{n}\end{matrix}f_{n}(k)\Omega_{n,\alpha\beta}(k).
\end{equation}

The structural evolution of bulk MnBi$_{2}$Te$_{4}$ is calculated under different pressures up to 9 GPa. The corresponding structural parameters are used to build  six-septuple-layer MnBi$_{2}$Te$_{4}$ under different pressures~\cite{Pressure}.

\begin{table}[t]
	\centering
	\renewcommand\arraystretch{2}
	\begin{tabular}{c|c|c|c|c|c|c}
		\hline\hline
		magnetic configuration & 1 & 2 & 3 & 4 & 5 & 6 \\\hline
		$\Delta E$ (meV) & 5.2 & 5.3 & 2.7 & 2.7 & 5.9 & 5.9 \\
		\hline\hline
	\end{tabular}
	\caption{The relative energy $\Delta E =E_{i}-E_{G}$ for different magnetic configurations of six-septuple-layer MnBi$_{2}$Te$_{4}$. The ground state magnetic configuration (A-type antiferromagnetic order) is taken as the reference (with its energy set as zero).}
	\label{tab1}
\end{table}

\section{\uppercase\expandafter{\romannumeral3}. Structural and Electronic Properties} 
Bulk MnBi$_{2}$Te$_{4}$ is an A-type antiferromagnetic layered material crystallized in $R\overline{3}m$ space group. There is a spatial inversion point at the center of the van der Waals gap. When the MnBi$_{2}$Te$_{4}$ is down to an even-septuple-layer film, this spatial inversion point (P) is still preserved (see Fig.~\ref{fig1}(a)). The antiferromagnetic order breaks the time-reversal symmetry, but the combined $\mathcal{PT}$ symmetry forces a vanishing integrated berry curvature within the first Brillouin zone. Several methods are used to remove the $\mathcal{PT}$ symmetry allowing the presence of QAHE, such as applying an external vertical electric field~\cite{ele} and tuning the stacking order~\cite{stack1,stack2}. Moreover, the $\mathcal{PT}$ symmetry can also be broken by tuning the magnetic configuration. Figure.~\ref{fig1}(c) shows the six compensated antiferromagnetic configurations breaking $\mathcal{PT}$ symmetry for six-septuple-layer MnBi$_{2}$Te$_{4}$, respectively. The time-reversed counterparts are not depicted. Table~\ref{tab1} lists the relative energy $\Delta E$ of the six compensated antiferromagnetic configurations to A-type antiferromagnetic ground state. The small magnitudes of $\Delta E$ imply that it is promising to realize these configurations by external manipulation, which we will discuss in the following part.

As shown in Figs.~\ref{fig2}(a)-(f), we calculate the band structures of six-septuple-layer MnBi$_{2}$Te$_{4}$ with the six compensated antiferromagnetic configurations in the presence of spin-orbit coupling, respectively. The band structure feature of MnBi$_{2}$Te$_{4}$ is still remained with band gaps equaling 45.7, 14.4, 14.9, 8.2, 6.5 and 9.5 meV. Figures.~\ref{fig2}(g)-(l) plot the corresponding orbital projected band structures around the $\Gamma$ point, illustrating that the band inversion between Bi-$p_{z}$ and Te-$p_{z}$ states still survives in all compensated antiferromagnetic configurations. Thus, these band gaps show the potential to harbor topologically nontrivial states. Then, by constructing a Wannier-based tight-binding model on Bi-$p_{z}$ and Te-$p_{z}$ orbitals, we calculate the anomalous Hall conductivity of all cases, as shown in Figs.~\ref{fig3}(a)-(f). Zero Hall plateaus are observed in the band gaps of m$_{1}$ and m$_{2}$ configurations, respectively. While quantized plateaus of $\sigma_{xy}=-e^{2}/h$ emerge, respectively, in band gaps of m$_{3}$, m$_{4}$, m$_{5}$, and m$_{6}$ configurations, indicating the presence of the QAHE with $\mathcal{C}=-1$. 

\begin{figure}
	\centering
	\includegraphics[width=0.48\textwidth]{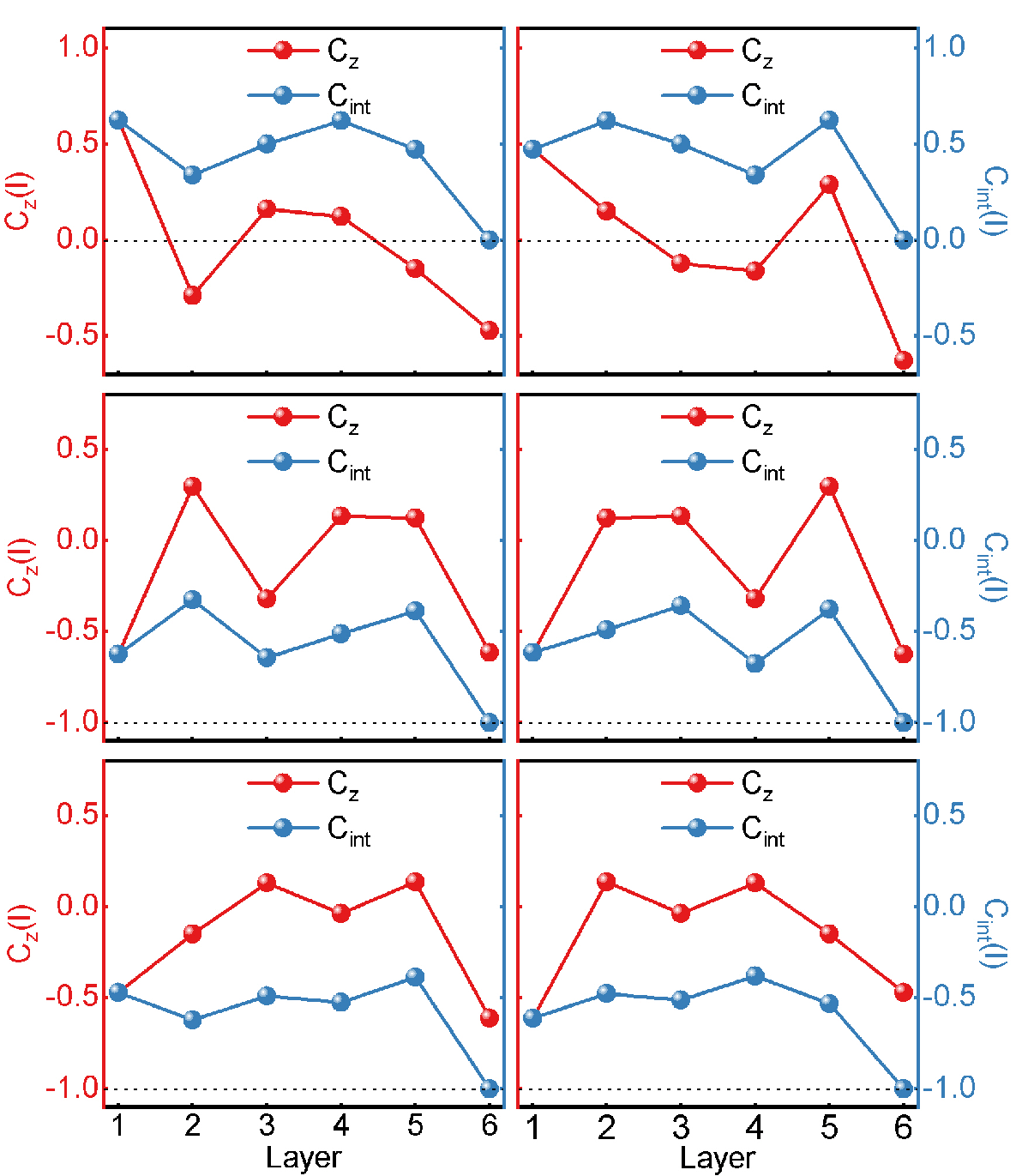}
	\caption{(a)-(f) Layer-resolved Chern number $\mathcal{C}_{z}(l)$ (red) and its summation $\mathcal{C}_{int}(l)$ (gray blue) from the first to the $l$th layer of six-septuple-layer MnBi$_{2}$Te$_{4}$ with m$_{1}$, m$_{2}$, m$_{3}$, m$_{4}$, m$_{5}$, and m$_{6}$ magnetic configurations, respectively.}
	\label{fig4}
\end{figure}

\section{\uppercase\expandafter{\romannumeral3}. Layer-resolved Chern Number} 
As a rule, the broken of $\mathcal{PT}$ symmetry will raise nonzero anomalous Hall conductivity in a MnBi$_{2}$Te$_{4}$ film, such as the layer Hall effect induced by a vertical electric field~\cite{layer-dep3}. However, the net anomalous Hall conductance is not observed in the m$_{1}$ and m$_{2}$ configurations. The layer-resolved Chern number is required to gain further insight into the emergence of different Chern numbers in this multilayer system. A tight-binding model of MnBi$_{2}$Te$_{4}$ is constructed on the basis of $\left\lbrace \left| \text{p}_\text{z,Bi}^+,\uparrow\right\rangle , \left| \text{p}_\text{z,Te}^-,\uparrow\right\rangle, \left| \text{p}_\text{z,Bi}^+,\downarrow\right\rangle , \left| \text{p}_\text{z,Te}^-,\downarrow\right\rangle \right\rbrace$~\cite{model1,model2}:
\begin{eqnarray}\label{model hamiltonian}
	\nonumber
	H = \sum_{i}[E_0+(-1)^{n_z}m_0] c_i^\dagger c_i + \sum_{\left\langle ij\right\rangle, \alpha}c_i^\dagger T_\alpha c_j + \text{H.c.}.
\end{eqnarray}
Here, $E_0 = \frac{1}{2}(M_0-\sum_{\alpha}B_\alpha)\sigma_0 \otimes \tau_z - \frac{1}{2}\sum_{\alpha}D_\alpha \sigma_0 \otimes \tau_0$, $T_{\alpha} =(B_\alpha \sigma_0 \otimes \tau_z + D_\alpha \sigma_0 \otimes \tau_0 - \text{i}A_\alpha \sigma_\alpha \otimes \tau_x)/2$ with $\alpha=x, y, z$, and $m_0 =m\sigma_z \otimes \tau_0$.  $\left\langle ij \right\rangle$ denotes the nearest neighboring coupling. $c_{i}^{\dagger}=(c_{i\uparrow}^{\dagger},c_{i\downarrow}^{\dagger})$ represents the creation operator for an electron at the $i$th site in spin degree of freedom. The first and third terms describe the bulk Hamiltonian of the topological insulator, while the second term determines the magnetic configurations. The orbital and spin Pauli matrices are represented by $\tau$ and $\sigma$, respectively. And the partial Chern number $\mathcal{C}_{z}(l)$ projected onto the $l$th layer can be expressed as:
\begin{eqnarray}\label{C_z(l)}
	\nonumber
	\mathcal{C}_{z}(l)=\frac{-4\pi}{A}Im\frac{1}{N_k}\sum_{k}\sum_{vv^{'}c}X_{vck}Y_{v^{'}ck}^{+}\rho_{v^{'}vk}(l).
\end{eqnarray}
Here, $N_{k}$ is the number of $k$ points in the first Brillouin zone, and $A$ is the area of a unit cell. $\psi_{vk}$ and $\psi_{ck}$ are the eigenstates of valence and conduction bands, respectively. $X=\left\langle \psi_{vk}|x|\psi_{ck}\right\rangle$ and $Y=\left\langle \psi_{vk}|y|\psi_{ck}\right\rangle$, respectively. $\rho_{v^{'}vk}(l)=\sum_{j\in l}\psi_{vk}^{*}(j)\psi_{v^{'}}k(j)$ represents the Bloch representation of the projection onto layer $l$ over orbitals $j$ in that layer. And we define its summation $\mathcal{C}_{int}(l)=\sum_{l}\mathcal{C}_{z}(l)$ from the first to the $l$th layer.

The calculated results are shown in Fig.~\ref{fig4}, which illustrates that the distribution of partial Chern numbers is inhomogeneous within this multilayer system. For m$_{1}$ and m$_{2}$ configurations, the $\mathcal{C}_{z}(l)$ is mainly distributed in the outermost layers with opposite signs, resulting in Chern number $\mathcal{C}=0$. In contrast, for the other four configurations, the $\mathcal{C}_{z}(l)$ in the outermost layers has same signs, resulting in Chern number $\mathcal{C}=-1$.

\section{\uppercase\expandafter{\romannumeral3}. The Effect of Pressure} 
\begin{figure}[t]
	\centering
	\includegraphics[width=0.48\textwidth]{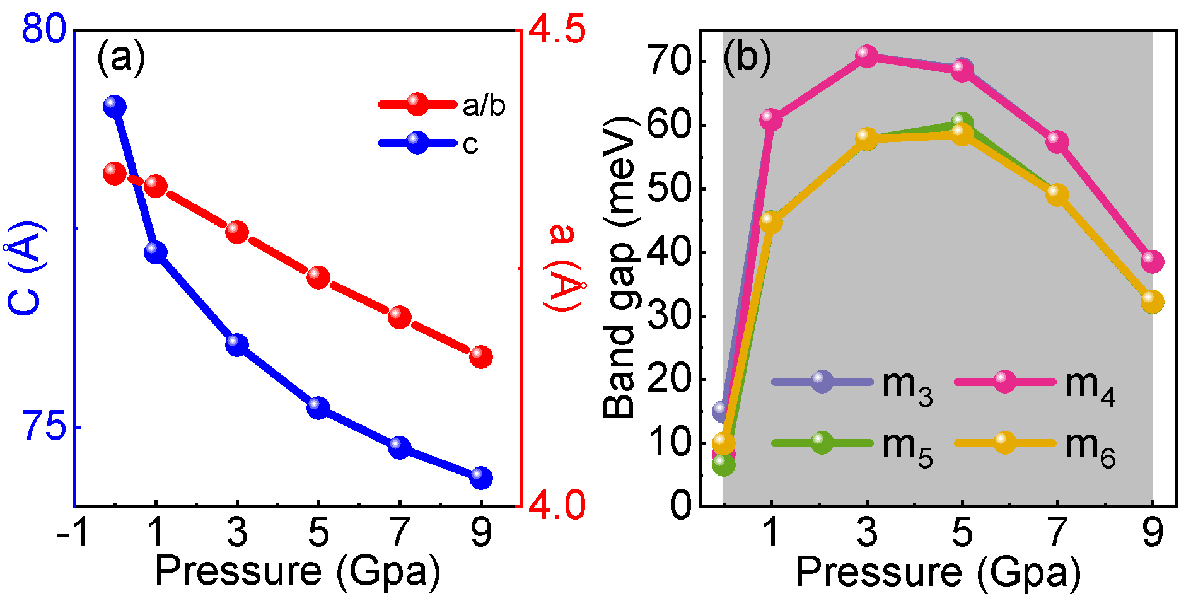}
	\caption{(a) The evolution of calculated in-plane and out-of-plane lattice parameters of MnBi$_{2}$Te$_{4}$ as a function of pressure. (b) The corresponding bulk band gaps of six-septuple-layer MnBi$_{2}$Te$_{4}$ with m$_{3}$, m$_{4}$, m$_{5}$, and m$_{6}$ magnetic configurations, respectively. The shadow area represents this region possessing QAHE with $\mathcal{C}=-1$.}
	\label{fig5}
\end{figure}

We have demonstrated that the QAHE can be realized in even-septuple-layer MnBi$_{2}$Te$_{4}$ with compensated antiferromagnetic orders. However, their topologically nontrivial gaps are lower than the room-temperature energy scale. The previous work has indicated that the magnitude of the inverted band gap is closely related to the strength of Te-Te quasicovalent bonding within the van der Waals gap~\cite{Te-Te}. High pressure is a clean method that can easily modulate the lattice parameters, especially for the out-of-plane lattice of two-dimensional materials~\cite{pre1,pre2}. Figure.~\ref{fig5}(a) shows the in-plane lattice parameter $a(b)$ and out-of-plane lattice parameter of six-septuple-layer MnBi$_{2}$Te$_{4}$ as a function of pressure up to 9 Gpa. We note that the out-of-plane lattice parameter decreases visibly implying the significant enhancement of interlayer interaction. As shown in Fig.~\ref{fig5}(b), the band gap increases strikingly in the range from 0 to 3 Gpa for m$_{3}$ and m$_{4}$ configurations, and decreases gradually in the following interval. For m$_{5}$ and m$_{6}$ configurations, the band gap keeps increasing in the range from 0 to 5 Gpa. It is exciting that, for these four configurations, the topologically nontrivial gaps all exceed the room-temperature energy scale once the pressure is applied, and their topological properties are always preserved with $\mathcal{C}=-1$ in the whole interval. The biggest band gap can reach 70.8 meV in m$_{3}$ and m$_{4}$ configurations at 3 Gpa. Thus, this system shows the potential to realize the high-temperature QAHE. 

\section{\uppercase\expandafter{\romannumeral4}. Coupling to the magnetic substrate}
\begin{figure}[t]
	\centering
	\includegraphics[width=0.48\textwidth]{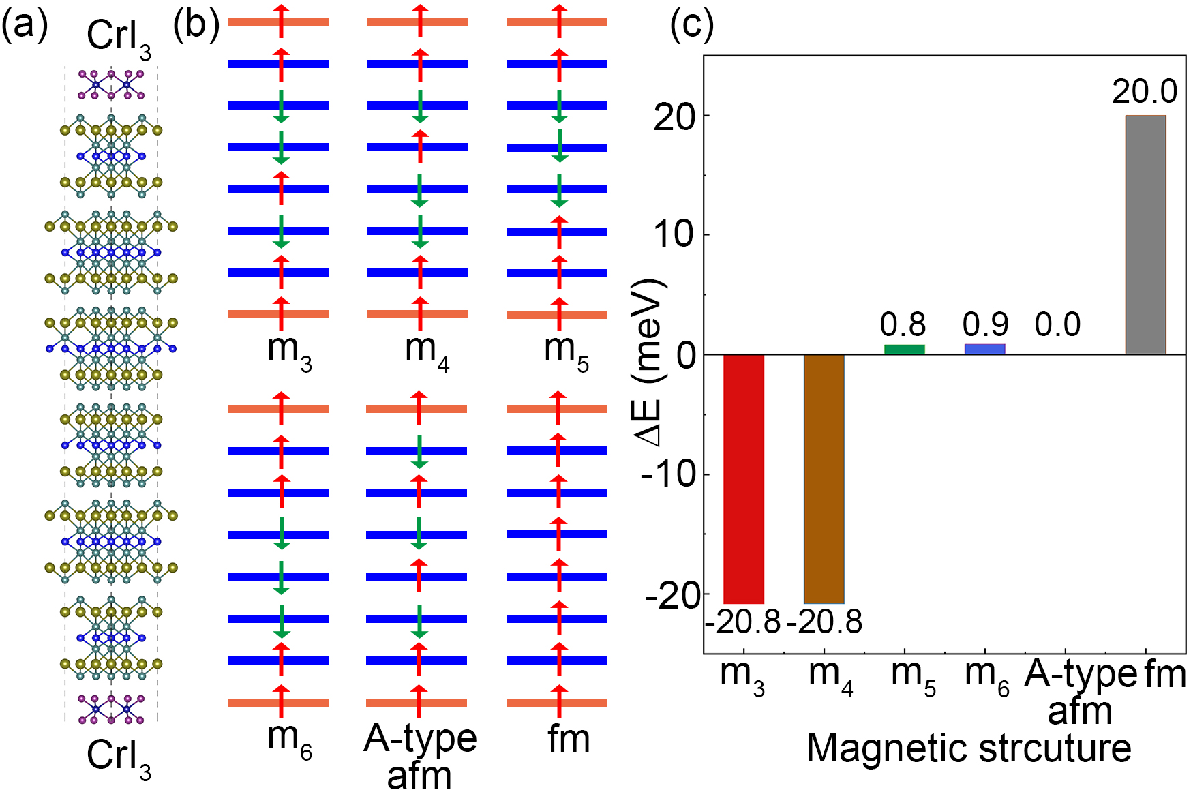}
	\caption{(a) Side view of MnBi$_{2}$Te$_{4}$/CrI$_{3}$ heterostructure, where the MnBi$_{2}$Te$_{4}$ is sandwiched by CrI$_{3}$. The spatial inversion symmetry is preserved. (b) Six different magnetic configurations of MnBi$_{2}$Te$_{4}$/CrI$_{3}$ heterostructure, i.e., MnBi$_{2}$Te$_{4}$ with m$_{3}$, m$_{4}$, m$_{5}$, m$_{6}$, A-type antiferromagnetic, and ferromagnetic magnetic configurations coupling to CrI$_{3}$ with magnetic moment along the $z$ direction. (c) The corresponding total energies $\Delta E$ relative to A-type antiferromagnetic configuration. }
	\label{fig6}
\end{figure}

The theoretical calculations and experimental measurement have illustrated that the magnetic configuration of MnBi$_{2}$Te$_{4}$ can be modulated by coupling to magnetic substrates, such as CrI$_{3}$~\cite{MBT1CrI3,MBT2CrI3}. The part as mentioned above lists that the relative energies of different compensated antiferromagnetic configurations to the A-type antiferromagnetic ground state are just around meV (See Table~\ref{tab1}). This stimulates us to explore the scheme to realize these compensated antiferromagnetic configurations in six-septuple-layer MnBi$_{2}$Te$_{4}$. The previous work shows that CrI$_{3}$ can not only provide a strong interface ferromagnetic coupling, but also preserve the topological properties of MnBi$_{2}$Te$_{4}$~\cite{MBT1CrI3}. Therefore, CrI$_{3}$ is used to construct the magnetic heterostructure where the six-septuple-layer MnBi$_{2}$Te$_{4}$ is sandwiched by CrI$_{3}$, as shown in Fig.~\ref{fig6}(a). Because the QAHE of compensated antiferromagnetic configurations requires that the magnetic moments of the outermost Mn2+ ions are parallel to each other, the magnetic moments in CrI$_{3}$ layer should also be set along the same direction, i.e., the $z$ or $-z$ direction. Figure~\ref{fig6}(b) presents six potential magnetic configurations of CrI$_{3}$/MnBi$_{2}$Te$_{4}$/CrI$_{3}$ heterostructure. The relative energies $\Delta E$ to the A-type antiferromagnetic configuration are depicted in Fig.~\ref{fig6}(c). Configurations m$_{3}$ and m$_{4}$ have negative $\Delta E$ manifesting that they are the most stable, i.e., they can be realized by coupling to magnetic substrates. While m$_{5}$, m$_{6}$, and ferromagnetic configurations have positive $\Delta E$ implying that they are unstable. Furthermore, the presence of magnetic substrates facilitates the reinforcement of the out-of-plane surface magnetization of MnBi$_{2}$Te$_{4}$. Therefore, we believe it is possible to realize the QAHE in compensated antiferromagnetic MnBi$_{2}$Te$_{4}$.

\section{\uppercase\expandafter{\romannumeral5}. Summary}
We theoretically demonstrate that the QAHE can be realized in compensated antiferromagnetic MnBi$_{2}$Te$_{4}$ by using the first-principles calculations. The model calculations give the layer-resolved Chern number in this multilayer system explaining the presence of different Chern numbers. Under the external pressure, the out-of-plane lattice parameter of MnBi$_{2}$Te$_{4}$ is compressed prominently leading to stronger Te-Te quasicovalent bonds. The topologically nontrivial gap exceeds the room-temperature energy scale in a wide range of pressure. In addition, we verify that the compensated antiferromagnetic configurations possessing the QAHE can be realized in MnBi$_{2}$Te$_{4}$/CrI$_{3}$ heterostructure. Our findings not only offer convincing proof of the presence of QAHE in compensated antiferromagnetic MnBi$_{2}$Te$_{4}$, but also provide a feasible scheme to obtain the corresponding magnetic configurations.

\section{ACKNOWLEDGMENTS}This work was supported by the National Natural Science Foundation of China (Grant Nos. 11974327 and 12004369), Anhui Initiative in Quantum Information Technologies (Grant No. AHY170000), China Postdoctoral Science Foundation (2023M733411 and 2023TQ0347), and the Innovation Program for Quantum Science and Technology (Grant No. 2021ZD0302800). We also thank the Supercomputing Center of University of Science and Technology of China for providing the high-performance computing resources.

\appendix


\begin{thebibliography}{99}
\bibitem{QAHEHaldane} F. D. M. Haldane, Model for a Quantum Hall Effect without Landau Levels, Phys. Rev. Lett. \textbf{61}, 2015 (1988).%
\bibitem{half-q} S.Q. Shen, Half quantized Hall effect, Coshare Science \textbf{02}, 01 (2024).
\bibitem{Cr-Bi2Se3} R. Yu, W. Zhang, H. J. Zhang, S. C. Zhang, X. Dai, and Z. Fang, Quantized Anomalous Hall Effect in Magnetic Topological Insulators, Science \textbf{329}, 61 (2010).%
\bibitem{Exp-Cr-Bi2Se3} C.-Z. Chang, J. S. Zhang, X. Feng, J. Shen, Z. C. Zhang, M.Guo, K. Li, Y. Ou, P. Wei, L.-L. Wang, Z.-Q. Ji, Y. Feng, S. H. Ji, X. Chen, J. F. Jia, X. Dai, Z. Fang, S.-C. Zhang, K. He, Y. Y. Wang, L. Lu, X.-C. Ma, and Q.-K. Xue, Experimental Observation of the Quantum Anomalous Hall Effect in a Magnetic Topological Insulator, Science \textbf{340}, 167 (2013).%
\bibitem{Rashba-Graphene} Z. Qiao, S. A. Yang, W.-X. Feng, W.-K. Tse, J. Ding, Y. G. Yao, J. Wang, and Q. Niu, Quantum anomalous Hall effect in graphene from Rashba and exchange effects, Phys. Rev. B \textbf{82}, 161414(R) (2010)%
\bibitem{2dTI} F. Liu, Two-dimensional topological insulators: past, present and future, Coshare Science \textbf{01}, 03 (2023).
\bibitem{TBG} M. Serlin, C. Tschirhart, H. Polshyn, Y. Zhang, J. Zhu, K. Watanabe, T. Taniguchi, L. Balents, and A. Young, Intrinsic quantized anomalous Hall effect in a moir\'{e} heterostructure, Science \textbf{367}, 900 (2020).%
\bibitem{TB-TMD} T. Li, S. Jiang, B. Shen, Y. Zhang, L. Li, Z. Tao, T. Devakul, K. Watanabe, T. Taniguchi, L. Fu, J. Shan, and K. F. Mak, Quantum anomalous Hall effect from intertwined moir\'{e} bands, Nature (London) \textbf{600}, 641 (2021).%
\bibitem{rhombo-gra} T. Han, Z. Lu, Y. Yao, J. Yang, J. Seo, C. Yoon, K. Watanabe, T. Taniguchi, L. Fu, F. Zhang, and L. Ju, Large quantum anomalous Hall effect in spin-orbit proximitized rhombohedral graphene, Science \textbf{384}, 647 (2024).%
\bibitem{mag-atom} Z. Li, Y. Han, and Z. Qiao, Chern number tunable quantum anomalous Hall effect in monolayer transitional metal oxides via manipulating magnetization orientation, Phys. Rev. Lett. \textbf{129}, 036801 (2022).%
\bibitem{e-e} A. L. Sharpe, E. J. Fox, A. W. Barnard, J. Finney, K. Watanabe, T. Taniguchi, M. A. Kastner, D. Goldhaber-Gordon, Emergent ferromagnetism near three-quarters filling in twisted bilayer graphene. Science \textbf{365}, 605 (2019).%
\bibitem{alter1} L. \v{S}mejkal, J. Sinova, and T. Jungwirth, Beyond conventional ferromagnetism and antiferromagnetism: A phase with nonrelativistic Spin and crystal rotation symmetry, Phys. Rev. X \textbf{12}, 031042 (2022).%
\bibitem{alter2} L. \v{S}mejkal, A.H. MacDonald, J. Sinova, S. Nakatsuji, and T. Jungwirth, Anomalous Hall antiferromagnets, Nat. Rev. Mater. \textbf{7}, 482 (2022).%
\bibitem{alqahe1} Y. Liu, J. Li and Q. Liu, Chern-Insulator Phase in Antiferromagnets, Nano Lett. \textbf{23}, 8650 (2023).%
\bibitem{alqahe2} P.-J. Guo, Z.-X. Liu, and Z.-Y. Lu, Quantum anomalous Hall effect in collinear antiferromagnetism, Npj Comput. Mater. \textbf{9}, 70 (2023).%
\bibitem{alqahe3} B. Wu, Y.-L. Song, W.-X. Ji, P.-J. Wang, S.-F. Zhang, and C.-W. Zhang, Quantum anomalous Hall effect in an antiferromagnetic monolayer of MoO, Phys. Rev. B \textbf{107}, 214419 (2023).
\bibitem{MBT-TI1} M. M. Otrokov, I. I. Klimovskikh, H. Bentmann, D. Estyunin, A. Zeugner, Z. S. Aliev, S. Gaß, A. Wolter, A. Koroleva, and A. M. Shikin et al., Prediction and observation of an antiferromagnetic topological insulator, Nature (London) \textbf{576}, 416 (2019).%
\bibitem{MBT-TI2} Y. Gong, J. W. Guo, J. H. Li, K. J. Zhu, M. H. Liao, X. Z. Liu, Q. H. Zhang, L. Gu, L. Tang, and X. Feng, Experimental realization of an intrinsic magnetic topological insulator, Chin. Phys. Lett. \textbf{36}, 076801 (2019).%
\bibitem{MBT-axion} D. Zhang, M. Shi, T. Zhu, D. Xing, H. Zhang, and J. Wang, Topological axion states in the magnetic insulator MnBi$_{2}$Te$_{4}$ with the quantized magnetoelectric effect, Phys. Rev. Lett. \textbf{122}, 206401 (2019).%
\bibitem{Weyl-MnBiTe} J. Li, Y. Li, S. Du, Z. Wang, B.-L. Gu, S.-C. Zhang, K. He, W. Duan, and Y. Xu,  Intrinsic magnetic topological insulators in van der Waals layered MnBi$_{2}$Te$_{4}$-family materials, Sci. Adv. \textbf{5}, eaaw5685 (2019).%
\bibitem{layer-dep1} M. M. Otrokov, I. P. Rusinov, M. Blanco-Rey, M. Hoffmann, A. Y. Vyazovskaya, S. V. Eremeev, A. Ernst, P. M. Echenique, A. Arnau, and E. V. Chulkov, Unique Thickness-Dependent Properties of the van der Waals Interlayer Antiferromagnet MnBi$_{2}$Te$_{4}$ Films, Phys. Rev. Lett. \textbf{122}, 107202 (2019).%
\bibitem{layer-dep2} S. Yang, X. Xu, Y. Zhu, R. Niu, C. Xu, Y. Peng, X. Cheng, X. Jia, Y. Huang, X. Xu, J. Lu, and Y. Ye, Odd-Even Layer-Number Effect and Layer-Dependent Magnetic Phase Diagrams in MnBi$_{2}$Te$_{4}$, Phys. Rev. X 11, 011003 (2021).%
\bibitem{layer-dep3} A. Gao, Y.-F. Liu, C. Hu, J.-X. Qiu, C. Tzschaschel, B. Ghosh, S.-C. Ho, D. B\'{e}rub\'{e}, R. Chen, and H. Sun et al., Layer Hall effect in a 2D topological axion antiferromagnet, Nature (London) 595, 521 (2021).%
\bibitem{layer-dep4} W. Liang, T. Hou, J. Zeng, Z. Liu, Y. Han, and Z. Qiao, Layer-dependent zero-line modes in antiferromagnetic topological insulators, Phys. Rev. B \textbf{107}, 075303 (2023).%
\bibitem{QAHE-MnBiTe} Y. Deng, Y. Yu, M. Z. Shi, J. Wang, X. H. Chen, and Y. Zhang, Quantum anomalous Hall effect in intrinsic magnetic topological insulator MnBi$_{2}$Te$_{4}$, Science \textbf{367}, 895 (2020).%
\bibitem{MBT1CrI3} H. Fu, C.-X. Liu, and B. Yan, Exchange bias and quantum anomalous Hall effect in the MnBi$_{2}$Te$_{4}$/CrI$_{3}$ heterostructure, Sci. Adv. \textbf{6}, eaaz0948 (2020).%
\bibitem{MBT2CrI3} Z. Ying, B. Chen, C. Li, B. Wei, Z. Dai, F. Guo, D. Pan, H. Zhang, D. Wu, X. Wang, S. Zhang, F. Fei, and F. Song, Large Exchange Bias Effect and Coverage-Dependent Interfacial Coupling in CrI$_{3}$/MnBi$_{2}$Te$_{4}$ van der Waals Heterostructures, Nano Lett. \textbf{23}, 765 (2023).%
\bibitem{mag-train1} S. K. Chong, Y. Cheng, H. Man, S. H. Lee, Y. Wang, B. Dai, M. Tanabe, T.-H. Yang, Z. Mao, and K. A. Moler et al., Intrinsic exchange biased anomalous Hall effect in an uncompensated antiferromagnet MnBi$_{2}$Te$_{4}$, Nat. Commun. \textbf{15}, 2881 (2024).%
\bibitem{mag-train2} B. Chen, X. Liu, Y.-H. Li, H. Tay, T. Taniguchi, K. Watanabe, M. H. W. Chan, J. Yan, F. Song, R. Cheng, and C.-Z. Chang, Even-odd layer-dependent exchange bias effect in MnBi$_{2}$Te$_{4}$ Chern insulator devices, Nano Lett. \textbf{24}, 8320 (2024).%
\bibitem{QAHE-c1} C. Lei, T. V. Trevisan, O. Heinonen, R. J. McQueeney, and A. H. MacDonald, Quantum anomalous Hall effect in perfectly compensated collinear antiferromagnetic thin films, Phys. Rev. B \textbf{106}, 195433 (2022).%
\bibitem{QAHE-c2} W. Liang, Z. Li, J. An, Y. Ren, Z. Qiao and Q. Niu, Chern Number Tunable Quantum Anomalous Hall Effect in Compensated Antiferromagnets, arXiv:2404.13305.%
\bibitem{vasp} G. Kresse and J. Furthm{\"u}ller, Efficient iterative schemes for ab initio total-energy calculations using a plane-wave basis set, Phys. Rev. B \textbf{54}, 11169 (1996).%
\bibitem{GGA} J. P. Perdew, K. Burke, and M. Ernzerhof, Generalized Gradient Approximation Made Simple, Phys. Rev. Lett. \textbf{77}, 3865 (1996).%
\bibitem{MP} H. J. Monkhorst and J. D. Pack, Special points for Brillouin-zone integrations, Phys. Rev. B \textbf{13}, 5188 (1976).%
\bibitem{GGA-U} L. Wang, T. Maxisch, and G. Ceder, Oxidation energies of transition metal oxides within the GGA+U framework, Phys. Rev. B \textbf{73}, 195107 (2006).%
\bibitem{wannier90} A. A. Mostofi, J. R. Yates, Y.-S. Lee, I. Souza, D. Vanderbilt, and N. Marzari, A tool for obtaining maximally-localised Wannier functions, Comput. Phys. Commun. \textbf{178}, 685 (2008).%
\bibitem{wannier90-QAHE} X. Wang, J. R. Yates, I. Souza, and D. Vanderbilt, Ab initio calculation of the anomalous Hall conductivity by Wannier interpolation, Phys. Rev. B \textbf{74}, 195118 (2006).%
\bibitem{Pressure} X. Fan, C.-H. Chang, W. T. Zheng, J.-L. Kuo, and D. J. Singh, The Electronic Properties of Single-Layer and Multilayer MoS$_{2}$ under High Pressure, J. Phys. Chem. C \textbf{119}, 10189 (2015).%
\bibitem{ele} S. Du, P. Tang, J. Li, Z. Lin, Y. Xu, W. Duan, and A. Rubio, Berry curvature engineering by gating two-dimensional antiferromagnets, Phys. Rev. Res. \textbf{2}, 022025(R) (2020).%
\bibitem{stack1} Y. Ren, S. Ke, W.-K. Lou, and K. Chang, Quantum phase transitions driven by sliding in bilayer MnBi$_{2}$Te$_{4}$, Phys. Rev. B \textbf{106}, 235302 (2022).%
\bibitem{stack2} Z. Niu, X.-L. Yu, D. Shao, X. Jing, D. Hou, X. Li, J. Sun, J. Shi, X. Fan, and T. Cao, Interlayer ferroelectric polarization modulated anomalous Hall effect in four-layer MnBi$_{2}$Te$_{4}$ antiferromagnets, Phys. Rev. B \textbf{109}, 174405 (2024).%
\bibitem{model1} H. Jiang, Z. Qiao, H. Liu, and Q. Niu, Quantum anomalous Hall effect with tunable Chern number in magnetic topological insulator film, Phys. Rev. B \textbf{85}, 045445 (2012).
\bibitem{model2} W. Liang, T. Hou, J. Zeng, Z. Liu, Y. Han, and Z. Qiao, Layer-dependent zero-line modes in antiferromagnetic topological insulators, Phys. Rev. B \textbf{107}, 075303 (2023).
\bibitem{Cz1} N. Varnava and D. Vanderbilt, Surfaces of axion insulators, Phys. Rev. B \textbf{98}, 245117 (2018).
\bibitem{Cz2} P. Deng, Y. Han, P. Zhang, S. K. Chong, Z. Qiao, and K. L. Wang, Tuning the number of chiral edge channels in a fixed quantum anomalous Hall system, Phys. Rev. B \textbf{109}, L201402 (2024).
\bibitem{Te-Te} R. Peng, Y. Ma, H. Wang, B. Huang, and Y. Dai, Stacking-dependent topological phase in bilayer MBi$_{2}$Te$_{4}$(M = Ge, Sn, Pb), Phys. Rev. B \textbf{101}, 115427 (2020).
\bibitem{pre1} X. Wang, Z. Li, M. Zhang, T. Hou, J. Zhao, L. Li, A. Rahman, Z. Xu, J. Gong, Z. Chi, R. Dai, Z. Wang, Z. Qiao, and Z. Zhang, Pressure-induced modification of the anomalous Hall effect in layered Fe$_{3}$GeTe$_{2}$, Phys. Rev. B \textbf{100}, 014407 (2019).
\bibitem{pre2} F. Yu, X. Zhu, X. Wen, Z. Gui, Z. Li, Y. Han, T. Wu, Z. Wang, Z. Xiang, Z. Qiao, J. Ying, and X. Chen, Pressure-induced dimensional crossover in a kagome superconductor, Phys. Rev. Lett. \textbf{128}, 077001 (2022).
\end{thebibliography}
\end{document}